\documentclass[a4paper]{llncs}
\usepackage[parfill]{parskip}    
\usepackage[english]{babel}
\usepackage[utf8]{inputenc}
\usepackage{amsmath}
\usepackage{amssymb}
\usepackage{graphicx}
\usepackage{tikzscale}
\usepackage{tikz}
\usepackage[normalem]{ulem}
\usetikzlibrary{positioning, arrows, shapes.misc, calc}

\usepackage{tabularx}

\usepackage{color}
\usepackage[colorinlistoftodos]{todonotes}

\newcommand{\guido}{}
\newcommand{\mic}{}

\newcommand{\true}{\mathtt{tt}}

\newcommand{\until}[2]{\mathbf{U}_{[#1,#2]}}
\newcommand{\eventually}[2]{\mathbf{F}_{[#1,#2]}}
\newcommand{\always}[2]{\mathbf{G}_{[#1,#2]}}

\usepackage{enumitem}
\newlist{indenteddesc}{description}{1}
\setlist[indenteddesc]{
  leftmargin=5em,  
  rightmargin=3em,
  labelindent=3em, 
  labelwidth=1.5em,
  labelsep=.5em
}

\newcommand{\JSD}[2]{{\rm JSD}[{#1}\parallel {#2}]}
\newcommand{\KL}[2]{{\rm KL}[{#1}\parallel {#2}]}

\usepackage{acronym}

\title{Property-driven State-Space Coarsening for Continuous Time Markov Chains \thanks{M.M., D.M.\ and G.S.\ are supported by the European Research Council under grant MLCS 306999. J.H.\ is supported by the EU project, QUANTICOL 600708.}}
\author{
Michalis Michaelides\inst1
\and Dimitrios Milios\inst1
\and\\ Jane Hillston\inst1
\and Guido Sanguinetti\inst1\inst2
}
\institute{School of Informatics, University of Edinburgh \and SynthSys, Centre for Synthetic and Systems Biology, University of Edinburgh}

\begin{document}

\acrodef{CTMC}{Continuous Time Markov Chain}
\acrodef{GP}{Gaussian Process}
\acrodef{MDS}{Multi-Dimensional Scaling}
\acrodef{DBSCAN}{Density-Based Spatial Clustering of Applications with Noise}
\acrodef{JSD}{Jensen-Shannon Divergence}

\maketitle

\begin{abstract} \label{abstract}

Dynamical systems with large state-spaces are often expensive to thoroughly explore experimentally. Coarse-graining methods aim to define simpler systems which are more amenable to analysis and exploration; most current methods, however, focus on a priori state aggregation based on similarities in transition rates, which is not necessarily reflected in similar behaviours at the level of trajectories. We propose a way to coarsen the state-space of a system which optimally preserves the satisfaction of a set of logical specifications about the system's trajectories. Our approach is based on Gaussian Process emulation and Multi-Dimensional Scaling, a dimensionality reduction technique which optimally preserves distances in non-Euclidean spaces. We show how to obtain low-dimensional visualisations of the system's state-space from the perspective of properties' satisfaction, and how to define macro-states which behave coherently with respect to the specifications. Our approach is illustrated on a non-trivial running example, showing promising performance and high computational efficiency.

\end{abstract}

\section{Introduction}
\label{introduction}

Reasoning about behavioural properties of dynamical systems is a central goal of formal modelling. Recent years have witnessed considerable progress in this direction, with the definition of formal languages \cite{danos2007rule,ciocchetta2009bio} and logics \cite{donze2010robust} which enable compact representations of dynamical systems, and mature reasoning tools to model-check properties in an exact \cite{kwiatkowska_prism_2011} or statistical way \cite{MC:YounesSimmons:INFCOMP2006:statMC,SB:Zuliani:2009:StatMC}.

While such advances are indubitably improving our understanding of dynamical systems, the applicability of these techniques in practical scenarios is still largely hindered by computational issues. In particular, systems with large state-spaces quickly become infeasible to analyse via exact methods due to the phenomenon of state-space explosion; even statistical methods may require computationally expensive and extensive simulations. 
State-space reduction methodologies aim to construct more compact representations for complex systems. Such reduced-state systems are generally amenable to more effective analysis and may yield deeper insights into the structure and dynamics of the system.

Broadly speaking, state-space reduction can be achieved by either model simplification, usually by abstracting some system behaviours into a simpler system, or state aggregation, often by exploiting symmetries or approximate invariances. A prime example of model simplification is the technique of time-scale separation, which replaces a large system with multiple weakly dependent sub-systems \cite{Bortolussi_CMSB2015}. Most aggregation methods, instead, are based on grouping different states with similar behaviour with respect to their transition probabilities. This idea is at the core of the concept of {\it approximate lumpability}, which extends the exact lumpability relationship by aggregating states based on a pre-defined metric on the outgoing exit rates \cite{deng2011optimal,BuchholzK14,AbateBCK15,TschaikowskiT15,Milios2015}.

In this paper we propose a novel state-space reduction paradigm by shifting the focus from the infinitesimal properties of states (i.e.\ their transition rates) to the global properties of trajectories. Namely, we seek to aggregate states that yield {\it behaviourally similar} trajectories according to a set of pre-defined logical specifications. Intuitively, two states will be aggregated if trajectories starting from either state exhibit similar probabilities of satisfying the logical specifications. We define a statistical algorithm based on statistical model checking and Gaussian Process emulation to define this behavioural similarity across the whole state-space of the system. We then propose a dimensionality reduction and clustering pipeline to aggregate states and define reduced (non-Markovian) dynamics. To illustrate our approach, we give a running example of model reduction for the Susceptible-Infected-Recovered-Susceptible (SIRS) model, a non-trivial, non-linear stochastic system widely used in epidemiology. Our results show that property-driven aggregation can yield an effective tool to reduce the complexity of stochastic dynamical systems, leading to non-trivial insights in the structure of their state-space.

\section{Background} \label{related-work}

\subsection{Population Continuous Time Markov Chains}

A \ac{CTMC} is a continuous-time Markovian stochastic process over a discrete state-space $\mathcal{S}$. We will consider only {\it population} models, where the state-space is organised along populations: in this case, the state-space is indexed by the counts of each population $n_i\in\mathbb{N}\cup \{0\}$. Population CTMCs (pCTMCs) are frequently used in many scientific and engineering domains; we will use here the notation of chemical reactions as it is widespread and intuitively appealing. Transitions in a pCTMC are denoted as 
\[r_1 X_1 + \ldots r_n X_n \xrightarrow{\tau(\vec{X})} s_1 X_1 + \ldots s_n X_n\]
meaning that $r_i$ particles of type $X_i$ are consumed and $s_j$ particles of type $X_j$ are produced when the specific transition takes place. $\tau(\vec{X})$ is a transition rate which depends on the current state of the system.

It is easy to show that waiting times between transitions are exponentially distributed random variables; this observation is the basis of exact simulation algorithms for pCTMCs, such as the celebrated Gillespie algorithm \cite{Gillespie1977}.
The Gillespie algorithm generates trajectories of a pCTMC by randomly choosing the next reaction to occur and the time to elapse until the reaction occurs.

\paragraph{Example 1.1}
We introduce here our running example, the Susceptible-Infected-Recovered-Susceptible (SIRS) model of epidemic spreading.
The SIRS model is a discrete stochastic model of disease spread in a population, where individuals in the population can be in one of three states, Susceptible, Infected and Recovered. There are different variations of the model, some open (individuals can enter and exit the system), others with individuals relapsing to a susceptible state after having recovered. Here, we consider a relapsing, closed system, which evolves in a discrete, 2-dimensional state-space, where dimensions are the number of Susceptible and Infected individuals in the population (Recovered numbers are uniquely determined since the total population is constant). We also introduce a spontaneous infection of a susceptible individual with constant rate, independent of the number of infected individuals, to eliminate absorbing states.

With a population size of $N$, states in the 2D space can be represented by $\vec{x}=(S, I), S \in \{0, \cdots, N\}, I \in \{0, \cdots, N-S\}$ for a total of $(N+1)(N+2)/2$ states. The chemical reactions for this system are:
\begin{indenteddesc}
 \item[{\bf infection}] $S + I \xrightarrow{\alpha} 2I$;
 \item[{\bf spontaneous infection}] $S \xrightarrow{\beta/5} I$;
 \item[{\bf recovery}] $I \xrightarrow{\beta} R$;
 \item[{\bf relapsing}] $R \xrightarrow{\beta} S$.
\end{indenteddesc}
We set the infection rate $\alpha = 0.005$, recovery rate $\beta = 0.01$, and population size $N = S+I+R = 100$, for a total of 5151 states in this SIRS system.
Sample trajectories of the system were simulated using the Gillespie algorithm.

\subsection{Temporal Logic and Model Checking}
We formally specify trajectory behaviours by using temporal logic properties. We are particularly interested in properties that can be verified on single trajectories, and assume metric bounds on the trajectories, so that they are observed only for a finite amount of time. Metric Interval Temporal logic (MITL) offers a convenient way to formalise such specifications.

Formally, MITL has the following grammar:
\[ \phi ::= \mathtt{tt}~|~\mu~|~\neg\phi~|~\phi_1\wedge\phi_2~|~\phi_1\until{T_1}{T_2}\phi_2,   \]
where $\true$ is the true formula, conjunction and negation are the standard boolean connectives, and the time-bounded until $\until{T_1}{T_2}$ is the only temporal modality.  Atomic propositions  $\mu$ 
are (non-linear) inequalities  on population variables. 
A MITL formula is interpreted over a function of time $\vec{x}$, and its satisfaction relation is  given as in \cite{MC:Maler:FORMATS2004:MITL}. 
More temporal modalities, such as the time-bounded eventually and always, can be defined in terms of the until operator: $\eventually{T_1}{T_2}\phi \equiv \true\until{T_1}{T_2} \phi$ and $\always{T_1}{T_2}\phi\equiv \neg \eventually{T_1}{T_2}\neg\phi$.

MITL formulae evaluate as true or false on individual trajectories; when trajectories are sampled from a stochastic process, the truth value of a MITL formula is a Bernoulli random variable. Computing the probability of such a random variable is a model checking problem. Model checking for MITL properties evaluated on trajectories from a CTMC requires the computation of transient probabilities; despite major computational efforts \cite{kwiatkowska_prism_2011}, this is seldom possible exactly due to state-space explosion. Statistical model checking (SMC) methods circumvent such problems by adopting a Monte Carlo perspective: by drawing repeatedly and independently sample trajectories, one may obtain an unbiased estimate of the truth probability, and statistical error bounds can be obtained by employing either frequentist or Bayesian statistical approaches \cite{MC:YounesSimmons:INFCOMP2006:statMC,SB:Zuliani:2009:StatMC}. It should be pointed out that such bounds do not carry the same guarantees as numerical results obtained say by transient analysis; however, simply by drawing more samples one may reduce the uncertainty in the bounds arbitrarily. 

\paragraph{Example 1.2}
MITL formulae can be used effectively to obtain behavioural characterisations of the system's trajectory. We turn again to the SIRS model to illustrate this concept.

Assume one may want to express a global bound on the virulence of the infection, so that the fraction of infected population never exceeds $\lambda$. This can be done by considering the formula $\phi_1$, defined as
\begin{equation}
\phi_1 ::= \always{0}{100} (I < \lambda N)
\label{eq:MITLphi1}\end{equation}
which translates to:
\[
\phi_1(\vec{x}) = \begin{cases} \mathtt{tt} &\mbox{if } I_{t} < \lambda N \text{ } \forall t \in [0, 100], \\ 
						     \neg\mathtt{tt} & \mbox{otherwise.}
				\end{cases}
\]
Statistical model checking of this formula is trivial: one simply draws a trajectory using Gillespie's algorithm, and monitors that the maximal number of infected does not exceed the specified threshold in the $[0,100]$ interval.

\section{Methodology}
\begin{figure}[h!tbp]
\centering	
\includegraphics[width=\textwidth]{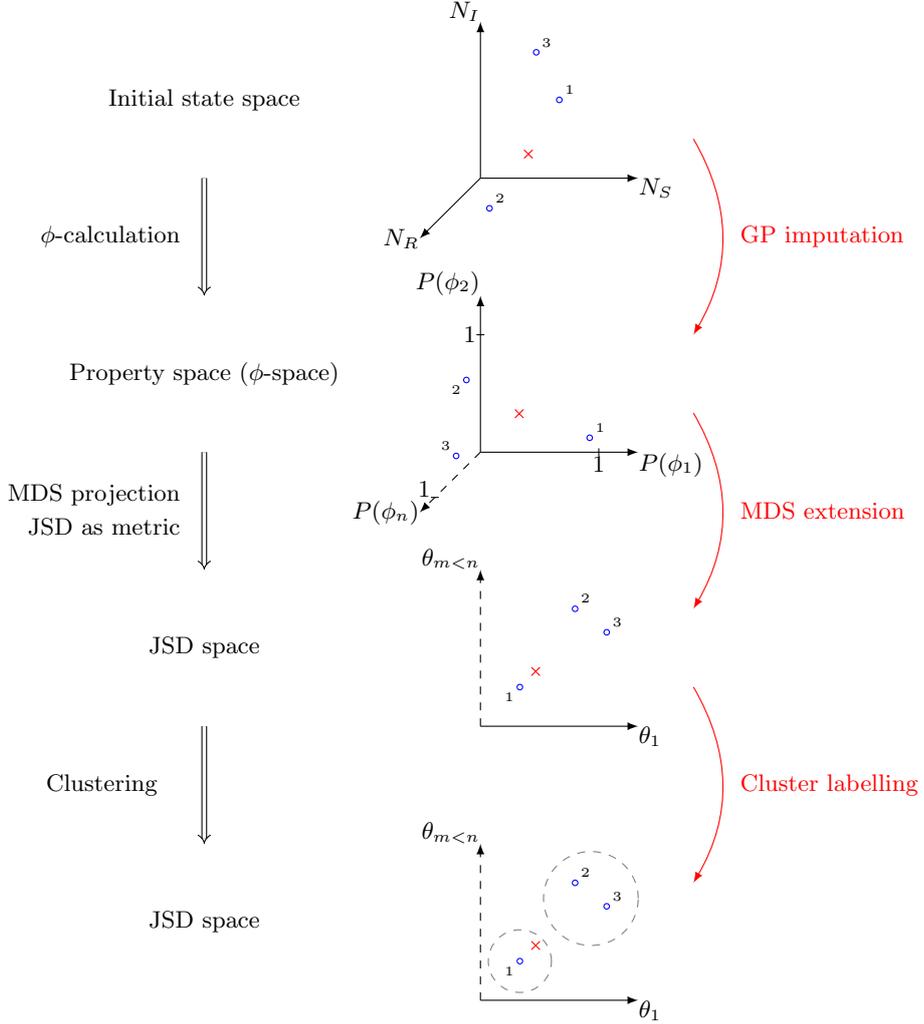}
\caption{The sequence of transformations from space to space are shown in the figure. States from the original state-space (blue circles 1-3) are projected to $\phi$-space according to satisfaction rate of set properties (found via simulation of the system). MDS is used to project from $\phi$-space to a space where \acs{JSD} of $\phi$ satisfaction probability distributions between states is preserved as Euclidean distance (in the figure, $\JSD{P_\phi(2)}{P_\phi(3)} < \JSD{P_\phi(1)}{P_\phi(2)}, \JSD{P_\phi(1)}{P_\phi(3)}$ so states 2, 3 are placed closer together than 1). The states are then clustered to produce macro-states. Out-of-sample states (red cross) can be projected to $\phi$-space, using GP imputation to estimate satisfaction probabilities. MDS extension allows projecting from $\phi$-space to \acs{JSD} space without moving the sampled states. The most likely cluster for the state to belong to (nearest centroid) is the macro-state it belongs to.}
\label{fig:spaces}
\end{figure}
\subsection{High level method description}

We first present a high-level description of the proposed methodology; the technical ingredients will be introduced in the following subsections. Figure~\ref{fig:spaces} provides an intuitive roadmap of the approach. The overarching idea is to provide a state-space aggregation algorithm which uses behavioural similarities as an aggregation criterion.

The input to the approach is a CTMC model and a set of MITL formulae $\phi_1,\ldots,\phi_n$ which define the behavioural traits we are interested in. \guido{We formalise some of the key concepts through the following definitions.}
\guido{
\begin{definition}
A \emph{coarsening map} $\mathcal{C}$ for a CTMC $\mathcal{M}$ is a surjective map 
\begin{equation}
\mathcal{M}: \mathcal{S} \xrightarrow{} \mathcal{R},
\label{eq:co-map}\end{equation}
from the state-space $\mathcal{S}$ of $\mathcal{M}$
to a finite set $\mathcal{R}$, such that $\mathrm{card}(\mathcal{S}) \geq \mathrm{card}(\mathcal{R})$.
\end{definition}}
\vspace{-0.5cm}
\guido{
\begin{definition}
The \emph{macro-states} of the coarsened system are the elements of the image of the coarsening map $\mathcal{C}$.
\end{definition}}
\guido{Therefore, the set of all macro-states is a partition of the set of initial states $\mathcal S$, where each element in the partition is a macro-state. In general, there is no way to retrieve the initial state configuration of the system only from information of the macro-state configuration, i.e., the coarsening entails an information loss.}

We illustrate the various steps of the proposed procedure in Figure \ref{fig:spaces}. The first step is to take a sample of possible initial states; we then evaluate the joint satisfaction of the $n$ formulae, given a particular state as initial condition. This implicitly defines a map \begin{equation}\Phi\colon\mathcal{S}\rightarrow[0,1]^{2^n}\label{property_Map}\end{equation}
which associates each initial state with the probability of each possible satisfaction pattern of the $n$ formulae. \guido{Notice that all of the $2^n$ possible truth values are needed to ensure correlations between properties are captured.} Constructing such a {\it property map} by exhaustive exploration of the state-space is clearly computationally infeasible; we therefore evaluate it (by SMC) on a subset of possible initial states, and then extend it using a statistical surrogate, a Gaussian Process (Figure \ref{fig:spaces} top).

The property representation contains the full information over the dependence of the properties of interest on the initial state. It can be endowed with an information-theoretic metric by using the \acs{JSD} between the resulting probability distributions. However, the high dimensionality and likely very non-trivial structure of the property representation may make this unwieldy. We therefore propose a dimensionality reduction strategy which maintains approximately the metric structure of the property representation using Multi-Dimensional Scaling (MDS; Figure \ref{fig:spaces} middle). MDS will also have the advantage of automatically identifying potentially redundant characterisations, as implied for example by logically dependent formulae.

The low-dimensional output of the MDS projection can then be visually inspected for groups of initial states ({\it macro-states}) with similar behaviours with respect to the properties. This operation is a {\it coarsening map}, which can also be automated by using a variety of clustering algorithms.

The model dynamics induce, in principle, a dynamics on this reduced space \mic{$\mathcal{R}$}. In practice, such dynamics will be non-Markovian and not easily expressible in a compact form; we propose a simple, simulation-based alternative definition which re-uses some of the computation performed in the previous steps to define an empirical, coarse-grained dynamics on the macro-states.

\subsection{Satisfaction probability as a function of initial conditions} \label{gaussian-processes}

The starting point for our approach consists of embedding the initial state-space into the property space, $\phi$-space. This is achieved by computing satisfaction probabilities for the $2^n$ possible truth patterns of the $n$ properties we consider. As in general these satisfaction probabilities can only be computed via SMC, this is potentially a tremendous computational bottleneck. To obviate this problem, we turn the computation of the property map into a machine learning problem: we evaluate the $2^n$ functions on a (sparse) subset of initial states, and predict their values on the remaining initial states using a \acf{GP}.

\acp{GP} have extensively been used in machine learning for regression purposes and it is in this context they are used here. A \ac{GP} is a generalisation of the multivariate normal distribution to function spaces with infinitely many dimensions; within a regression context, \acp{GP} are used to provide a flexible prior distribution over the set of candidate functions underpinning the hypothesised input-output relationship. Given a number of input-output observations (training set), one can use Bayes's rule to condition the \ac{GP} on the training set, obtaining a posterior distribution over the regression function at other input points. For a review of \acp{GP} and their uses in machine learning, we refer the reader to \cite{Rasmussen:Gaussian06}.

In our setting, the input-output relationship is the property map from initial states to satisfaction probabilities of the properties. This function is defined over a discrete space, but we can use the population structure of the pCTMC to embed the state-space $\mathcal{S}$ in a (subset) of $\mathbb{R}^D$ for some $D$. We can then treat the problem as a standard regression problem, learning a function 
$f_{\phi}\colon\mathbb{R}^D\rightarrow\mathbb{R}^{2^n}$. 

\paragraph{Remark.}
\acp{GP} have already been used to explore the dependence of the satisfaction probability of a formula on model parameters in the so-called Smoothed Model Checking approach \cite{smoothedMC}. There, the authors proved a smoothness result which justified the use of smoothness-inducing \acp{GP} for the problem. It is easy to see that such smoothness does not hold in general for the function $f_{\phi}$; for example, the probability of satisfying the formula $x(0)>N$ has a discontinuity at $x=N$. \guido{However, since we only ever evaluate $f_{\phi}$ on a discrete set of points, the lack of smoothness is not an issue, as a continuous function can approximate arbitrarily well a discontinuous function when restricted to a discrete set.}

\paragraph{Example 1.3} 
We exemplify this procedure on the SIRS example. We consider here three properties of interest: the global bound encoded in formula $\phi_1$ defined in equation \eqref{eq:MITLphi1}, and two further properties encoded as 
\begin{align}
\phi_2 &::=  \eventually{0}{60} \always{0}{40} (0.05N \leq I \leq 0.2N),\label{eq:MITLphi2} \\
\phi_3 &::=  \eventually{30}{50} (I > 0.3N).\label{eq:MITLphi3}
\end{align}

Satisfaction of $\phi_2$ requires that the infection has remained within 5 to 20\% of the total population for 40 consecutive time units, starting anytime in the first 60 time units; satisfaction of $\phi_3$ requires that the infection peaks at above 30\% between time 30 and time 50.

The property map in this case would have an 8-dimensional co-domain, representing the probability of satisfaction for each of the $2^3$ possible truth values of the three formulae. Figure~\ref{fig:phi_space} plots the probability of satisfaction for the three formulae individually, as we vary the initial state. In this case, 10\% of all possible initial states were randomly selected and numerically mapped to the property space via SMC, while the satisfaction probabilities for the remaining 90\% were imputed using GPs. We see that throughout most of the state-space the second property has low probability. Also it is of interest to observe the strong anti-correlation between the first and third properties: intuitively, if there is very high probability that the infection will be globally bounded below 40\% of individuals, it becomes more difficult to reach a peak at above 30\%.

\subsection{Dimensionality reduction of behaviours}\label{multi-dimensional-scaling-mds}

Once states are mapped onto $\phi$-space, reducing dimensionality of this space is useful to remove correlations and redundancies in the properties tracked. Properties may often capture similar behaviour, leading to strong correlations in their satisfaction probability.
Reducing the dimensionality of the property space mostly retains the information of how behaviour differs from state to state, eliminating redundancies. Moreover, reduced dimensional mappings can aid practitioners to visually identify structures within the state-space of the system.

In order to quantify the similarity of different initial states with respect to property satisfaction, the \ac{JSD} between the probability distributions of property satisfaction is used as a metric. \ac{JSD} is an information theoretic symmetric distance between probability distributions --- the higher the difference between the distributions, the higher \ac{JSD} is. Between two distributions, $P, Q$, \ac{JSD} is defined as
\[
\JSD{P}{Q} = \frac{1}{2} (\KL{P}{M} + \KL{Q}{M}),
\]
where $M = 0.5(P+Q)$ the average of the distributions, and $\KL{P}{Q}=\sum_i P(i) \log\frac{P(i)}{Q(i)}$, the Kullback-Leibler divergence. 

The JSD enables us to derive a matrix of pairwise distances in property space between different initial states. Such a distance is not Euclidean, and is defined in the high-dimensional property space. To map the initial states in a more convenient, low-dimensional space, we employ a dimensionality reduction technique known as Multi-Dimensional Scaling (MDS) \cite{BorgGroenen2005}.

MDS has its roots in the social science literature; it is a valuable and widely used tool in psychology and similar fields where data is collected by assessing similarity between pairs.


Given some points $X$ in an $m$-dimensional space, MDS finds the position of corresponding points $Z$ in an $n$-dimensional space, where usually $n<m$, such that a given metric between points is optimally preserved. In the most common case, (also known as Torgerson–Gower scaling or Principal Component Analysis), the metric to be preserved is the Euclidean distance, and is preserved by minimisation of a loss function. This function is generally known as {\it stress} for metric MDS, but specifically for classical MDS as {\it strain}.

For the classical MDS case, the projection is achieved by eigenvalue decomposition of a distance matrix of the (normalised) points $XX^\top$, and subsequently reconstructing the points from the $n$ largest (eigenvector, eigenvalue) pairs. This results in $Z$, a projection of the points to an $n$-dimensional space, where Euclidean distance is optimally preserved.

In the classical MDS definition, the MDS projection is defined statically for the available data points, and needs ab initio re-computation if new points become available. In \cite{bengio2004out}, the method is extended to new points by constructing a new dissimilarity matrix of new points to old ones, by which the projection of new points will be consistent to that of the old points. The kernel for this new matrix achieves this by replacing the means required for centring with expectations over the old points; such that for points $x, y \in X$
\[
\tilde{K}(x, y) = -\frac{1}{2} \bigg( d^2(x, y) - \frac{1}{n} \sum_{x'} d^2(x', y) - \frac{1}{n} \sum_{y'} d^2(x, y') + \frac{1}{n^2} \sum_{x', y'}d^2(x', y')] \bigg),
\]
where $\tilde{K}(x, y)$ is the kernel used for the dissimilarity matrix, is replaced by 
\[
\tilde{K}(a, b) = -\frac{1}{2} \bigg( d^2(a, b) - E_x[d^2(x, a)] - E_{x'}[d^2(b, x')] + E_{x, x'}[d^2(x, x')] \bigg),
\]
where $a$ can be an out-of-sample point ($a \notin X, b\in X$).

This reconstructs the dissimilarity matrix for the original points exactly, and allows us to generalise to out-of-sample points and find their positions in the embedding learned, as described in \cite{bengio2004out}. Extending MDS allows us to create macro-states based on samples of points, and then project new points on the space created by MDS to find in which clusters they belong.

\begin{figure}[h!tbp]
\begin{center}
\includegraphics[height=1.7in]{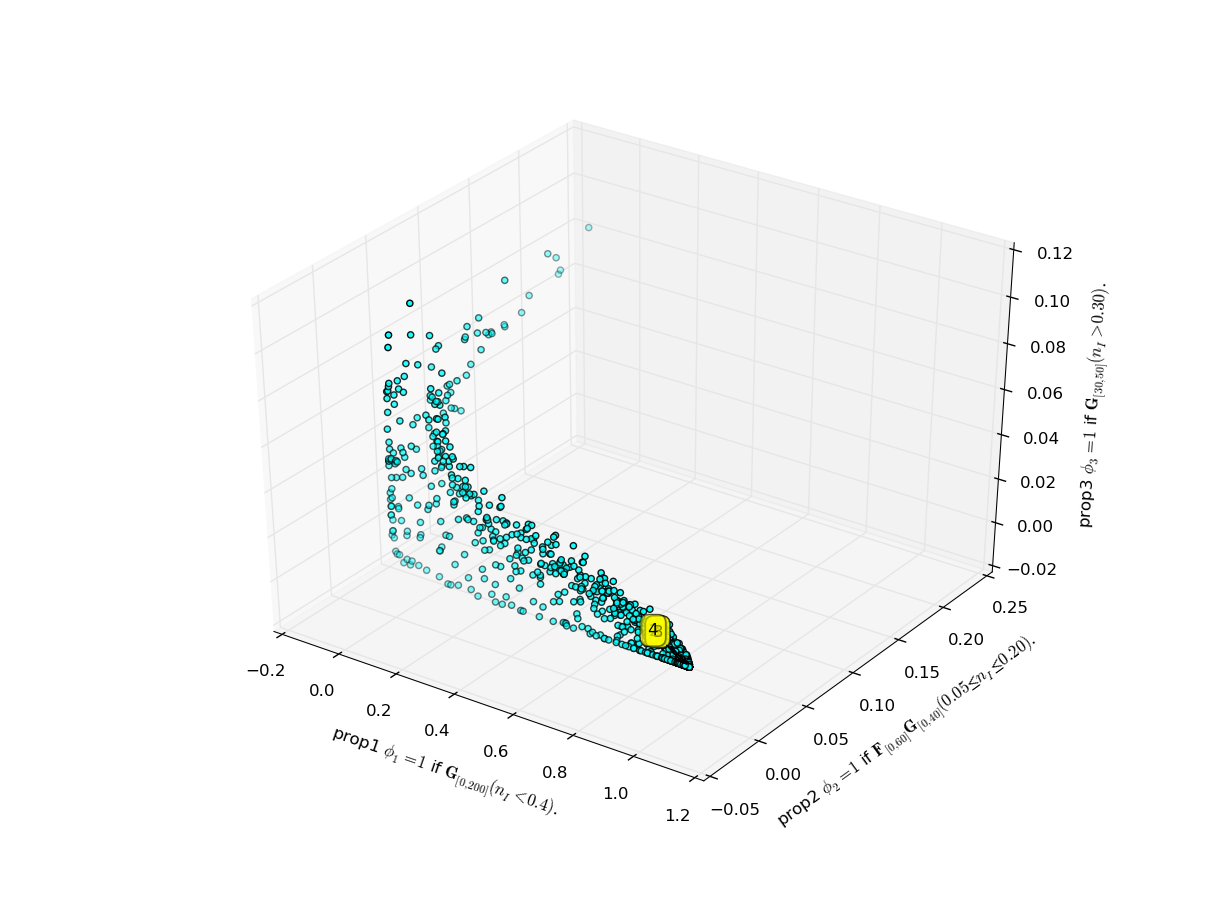}
\includegraphics[height=1.7in]{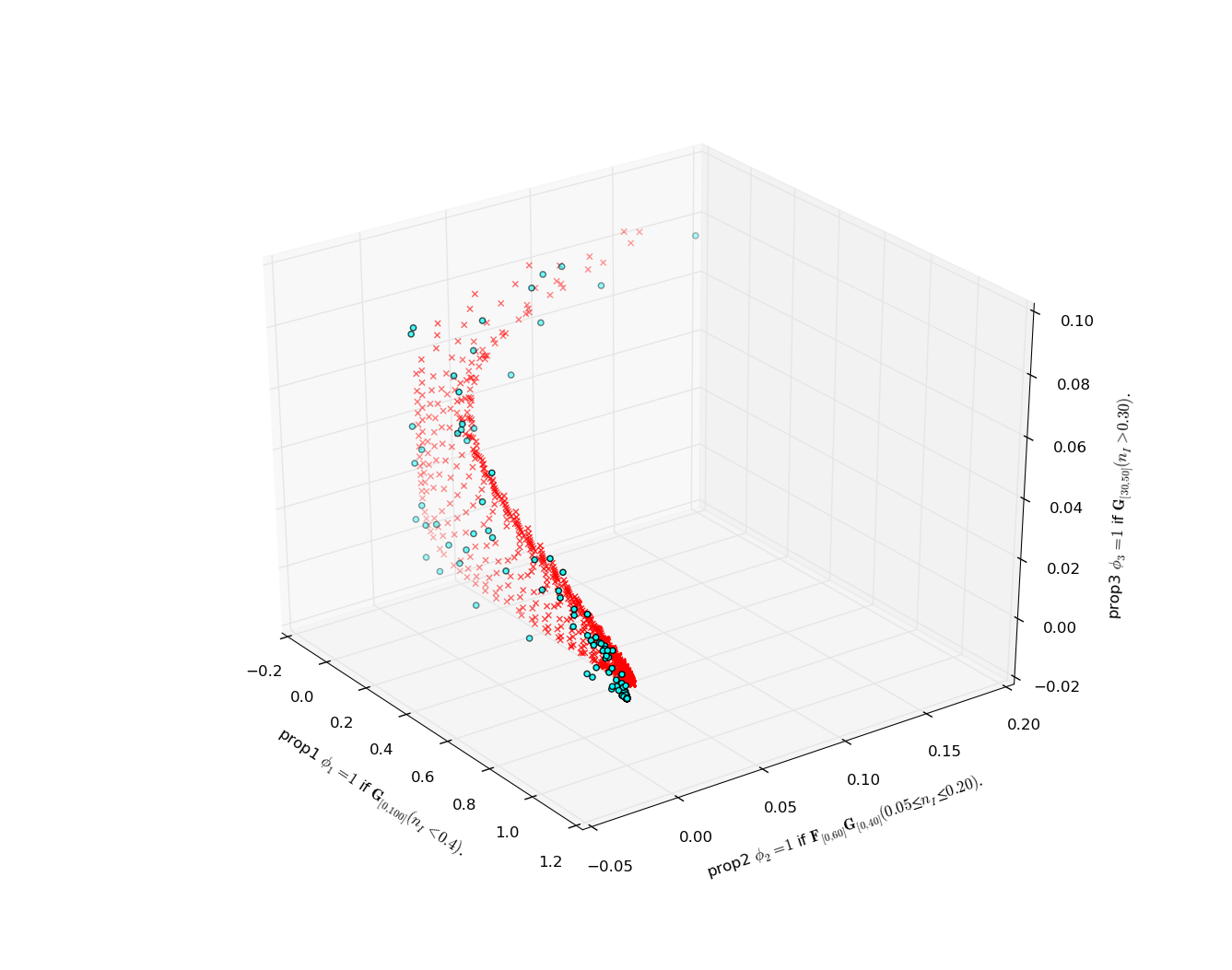}
\caption{Left: Projection of states in $\phi$-space via SMC (trajectory simulations for each initial state). Notice the non-trivial state distribution structure. Right: Projection of states in $\phi$-space using SMC for 10\% of the states, and GP regression to estimate $P(\phi)$ for the rest 90\% of states (red crosses).}
\label{fig:phi_space}
\end{center}
\end{figure}

\paragraph{Example 1.4}
We have introduced three properties in Equations (\ref{eq:MITLphi1}), (\ref{eq:MITLphi2}) and (\ref{eq:MITLphi3}), and the associated property map. This has an eight-dimensional co-domain, but already some of its properties can be gleaned by the three-dimensional plot of the single-formula probabilities shown in Figure~\ref{fig:phi_space}. Particularly, these reveal strong negative correlations, indicating that MDS may prove fruitful.


%
%
%

\begin{figure}[h!tbp]
\begin{center}
\includegraphics[height=1.7in]{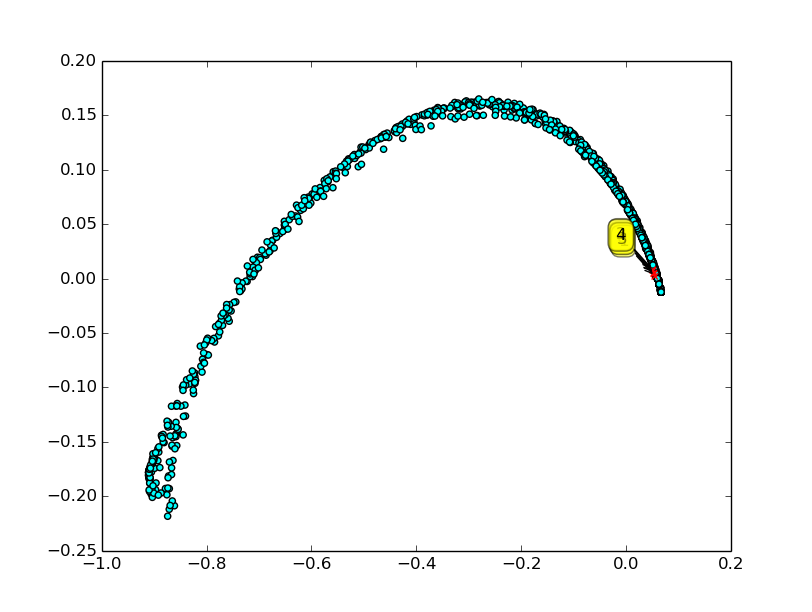}
\includegraphics[height=1.7in]{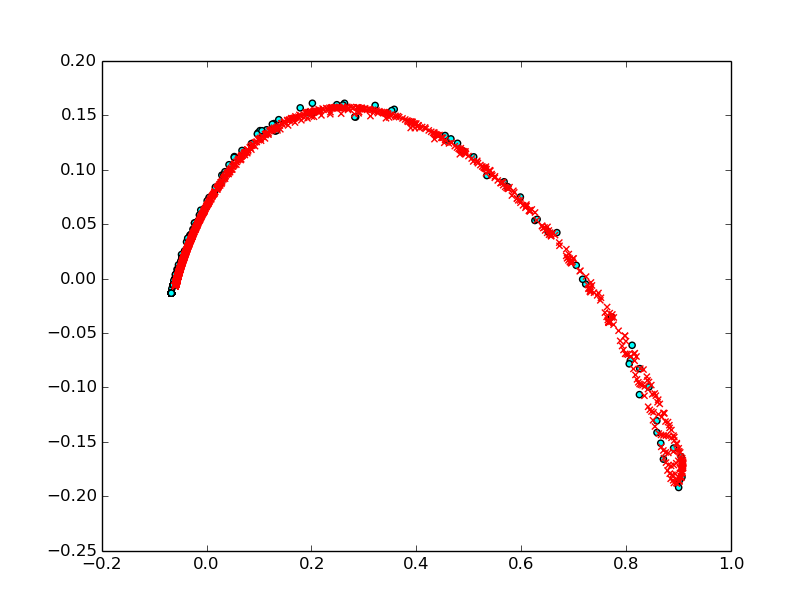}
\caption{Left: $P(\phi_1, \phi_2, \phi_3)$ estimated via SMC for each state. MDS was then used to project them from an 8D to a 2D space. Right: GP estimates of $P(\phi_1, \phi_2, \phi_3)$ for 90\% of states (red crosses) produce an almost identical MDS projection.}
\label{fig:MDS_space}
\end{center}
\end{figure}

Figure \ref{fig:MDS_space} shows the states projected to a 2D space were proximity implies similar probability distribution over property satisfaction. This was achieved using MDS to project the states, with \ac{JSD} used as the metric to be preserved as Euclidean distance in the new 2D space. Elements of the square-shaped structure visible in $\phi$-space (figure \ref{fig:phi_space}) are preserved, with the subset of states giving rise to higher probabilities for property $\phi_2$ (top of Figure \ref{fig:phi_space}) appearing further from the connected outline (bottom left group in Figure \ref{fig:MDS_space}).

\subsection{Clustering and structure discovery}\label{clustering-and-structure-discovery}

\begin{figure}[htbp]
\centering
\includegraphics[height=2in]{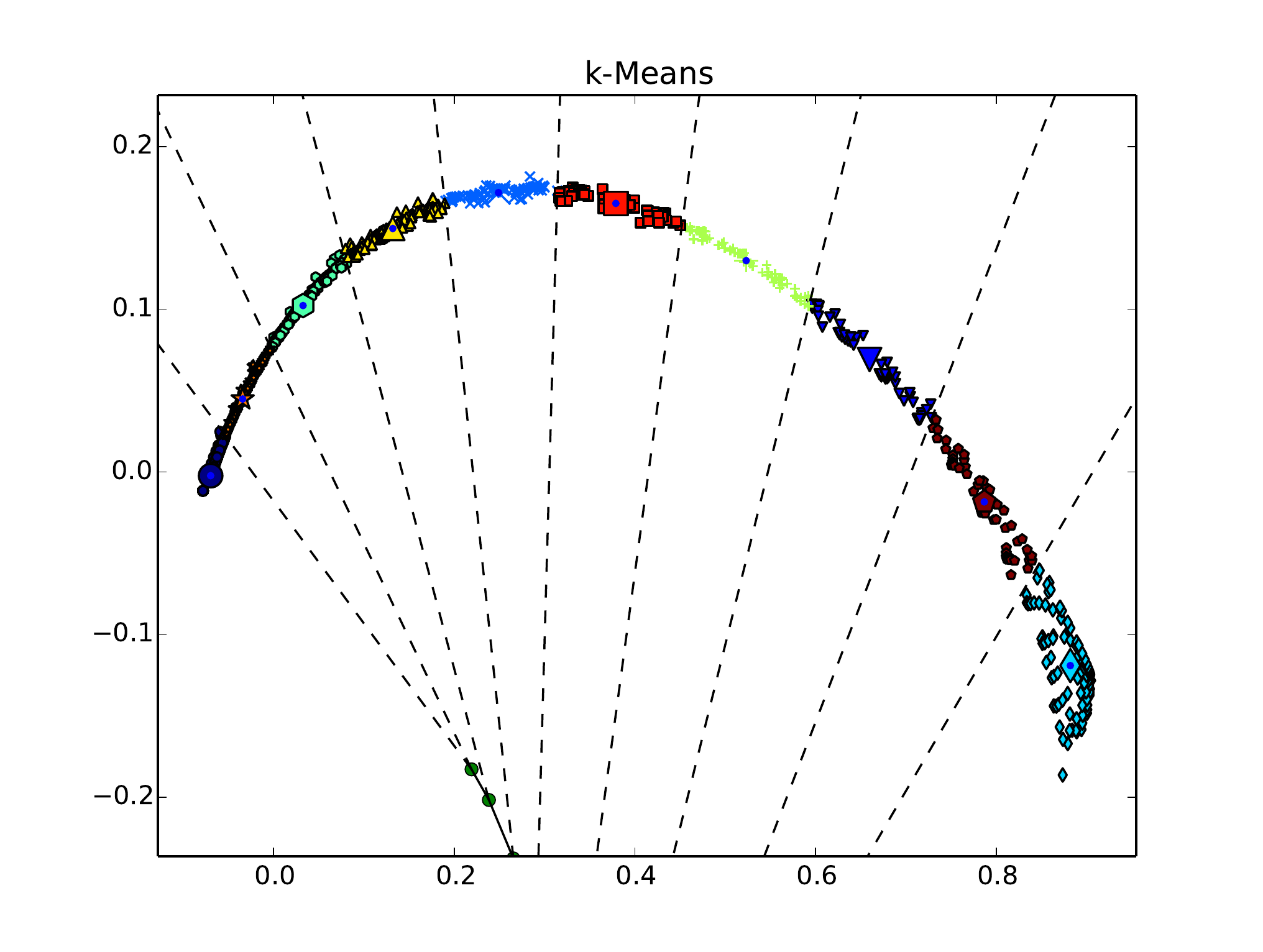}
\caption{The states were clustered in the space created by the MDS projection and coloured accordingly, using $k$-means (10 clusters). Since the Euclidean distance in this space is representative of distance in probability distributions over properties, states with different behaviour should be in different clusters.}
\label{fig:clusters}
\end{figure}

The MDS projection enables us to visually appreciate the existence of non-trivial structures within the state-space, such as clusters of initial states that produce similar behaviours with respect to the property specification.
Our intuition is that such structures should form the basis to define macro-states of the system, groups of states that will exhibit similar satisfaction probabilities for the properties defined. To automate this process, we propose to use a clustering algorithm to define macro-states. Since our goal is to group states with similar behaviours, we adopt $k$-means clustering \cite{Bishop2006}, which is based on the Euclidean distance of the states in the MDS space (representative of the \ac{JSD} between the probability satisfaction distributions). $k$-means requires specification of the desired number of clusters (the $k$ parameter); this allows the user to select the level of coarsening required. Figure~\ref{fig:clusters} shows the clusters produced in the reduced MDS space for the running SIRS model example, where we set the number of clusters $k=10$.

\subsection{Constructing coarse dynamics}\label{coarse-dynamics}

Once states have been grouped into macro-states, a major question is how to construct dynamics for the now coarsened system. The coarsened system naturally inherits dynamics from the original (fine-grained) system; however, such dynamics are non-Markovian, and in general fully history dependent so that transition probabilities would have the form 
\begin{equation}
p(k', t \vert k, h) = p(k'|k,t, h) p(t|k,h),
\label{nonMark}\end{equation}
where $h$ denotes the history of the process.
Simulating such a non-Markovian system is very difficult and likely to be much more computationally expensive than simulating the original system.

We therefore seek to define approximate dynamics which are amenable to efficient simulation, but still capture aspects of the non-Markovian dynamics. The most natural approximation is to replace the system with a semi-Markov system: transitions are still history-independent, but the distribution of sojourn times is non-exponential. To evaluate the sojourn-time distribution, we resort to an empirical strategy, and construct an empirical distribution of sojourn times by re-using the simulated trajectories of the fine system that were drawn to define the coarsening. In other words, once a clustering is defined, we retrospectively inspect the trajectories to construct a histogram distribution of sojourn times, approximating $p(t|k)$.

A possible drawback of this semi-Markov approximation is that it may introduce transitions which are actually impossible in the original state-space. This is because states were clustered based on behaviour rather than transition rates, and therefore states that are actually quite far in the original state-space may end up being clustered together. Since the identity of the original states is lost after the coarse graining, impossible transitions may be introduced.

Retrospectively inspecting whole system trajectories, rather than agnostically examining cluster transitions of the original system with a uniform initial state distribution within the cluster, ameliorates this problem.
Similarly, estimates of $p(k'|t,k)$ are produced from the same trajectories; these are the macro-state transition frequencies in each bin of the sojourn time probability histogram. This method avoids a lot of impossible trajectories one might generate, if the above probabilities were estimated by sampling randomly from initial states in a macro-state and looking at when the macro-state is exited and to which macro-state the system transitions. Assuming the original system has a steady state, the empirical dynamics constructed here capture this steady state macro-state distribution; however, accuracy of transient dynamics suffers, and the coarsened system enters the steady state faster than the original system.

\begin{figure}[h!tbp]
\centering
\includegraphics[height=2.8in]{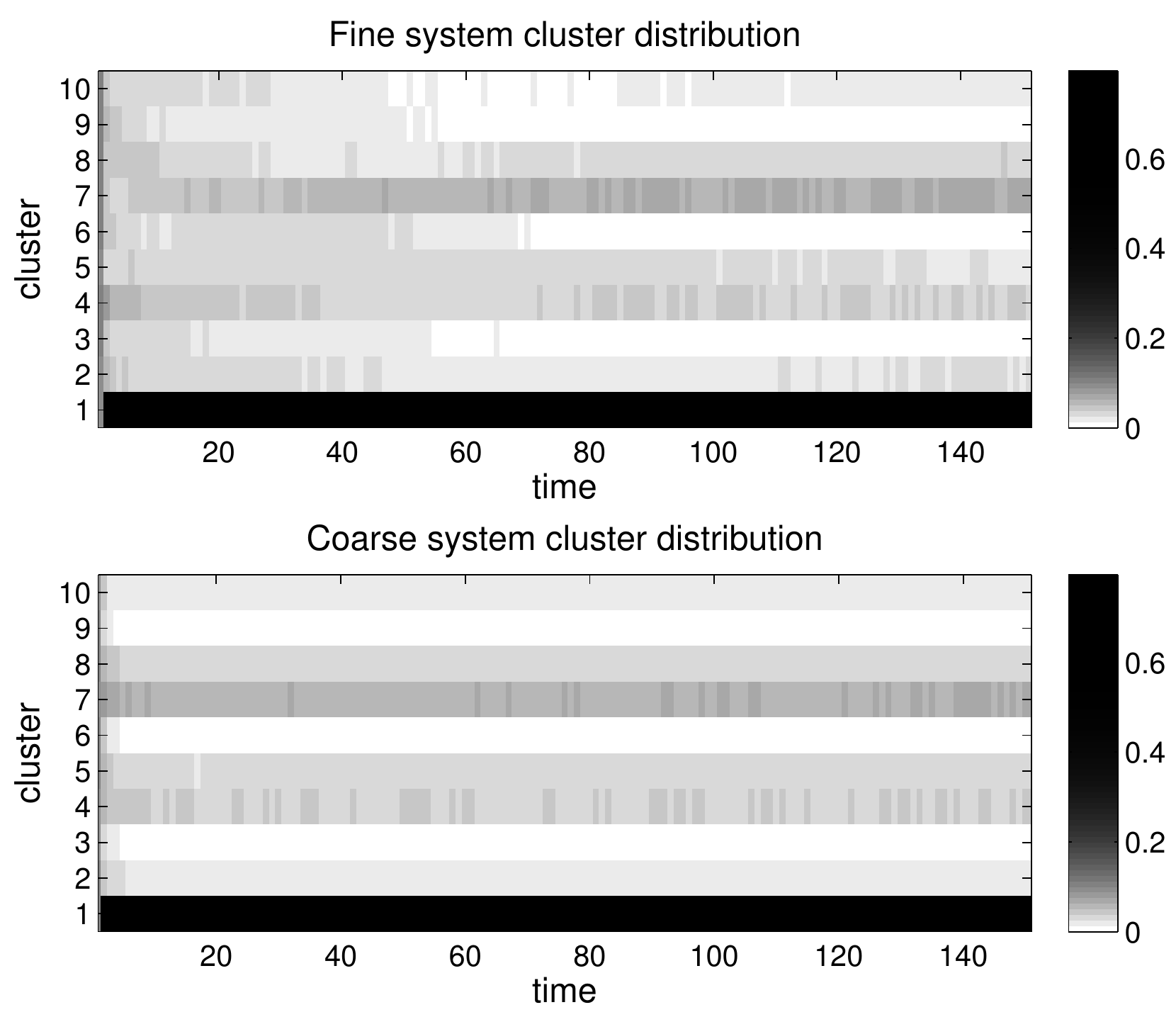}
\caption{Evolution of the macro-state histograms over time}
\label{fig:distributions}
\vspace{-0.6cm}
\end{figure}

\paragraph{Example 1.5}
We illustrate and evaluate the quality of the coarsened trajectories with respect to the original ones on the SIRS example. In particular, we examine the probability distribution over the macro-states at different times in the evolution of the system.
The macro-state distribution has been estimated empirically by sampling trajectories using the Gillespie algorithm for the fine system, and our coarse simulation scheme for the coarsened system.
We have then constructed histograms to capture the distribution of the categorical random variables that represent the macro-state.
Finally, we measure the histogram distance between histograms obtained from the fine and the coarse systems.
Figure \ref{fig:distributions} depicts the evolution of the macro-state histograms over time.

\begin{figure}[h!tbp]
\centering
\includegraphics[height=1.6in]{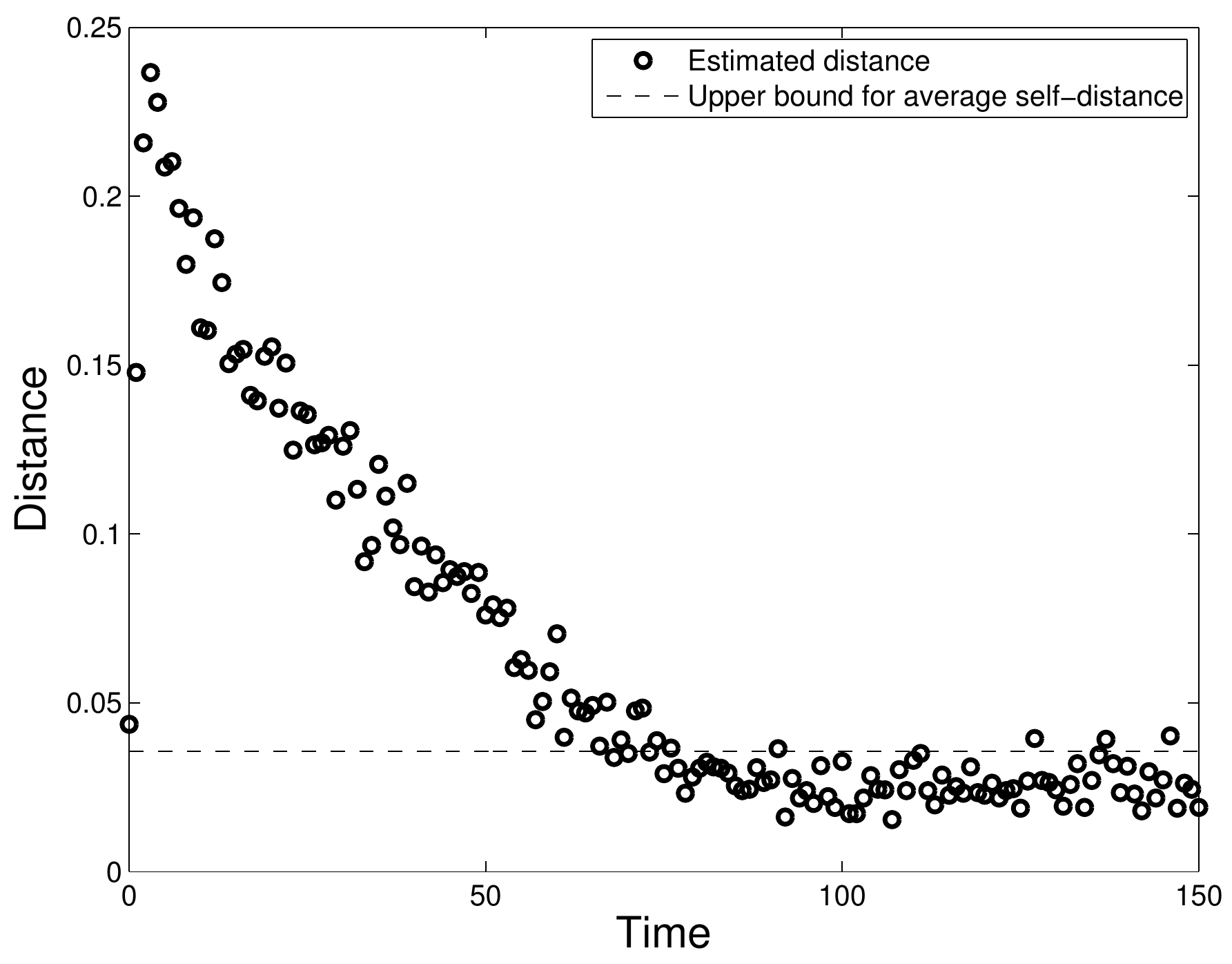}
\caption{Evolution of the macro-state histogram distances over time}
\label{fig:distances}
\vspace{-0.6cm}
\end{figure}

\paragraph{Quality of Approximation}
In order to put any distance between empirical distributions into context, this has to be compared with the corresponding average self-distance, which is the expected distance value when we compare two samples from the same distribution.
In this work, we estimate the self-distance using the result of \cite{Cao2006}: given $N$ samples and $K$ bins in the histogram, an upper bound for the average histogram self-distance is given by $\sqrt{(4 K)/(\pi N)}$.
In our example, we have $K = 10$ histogram bins, which are as many as the macro-states.
In practice, a distance value smaller than the self-distance implies that the distributions compared are virtually identical for a given number of samples.
In Figure \ref{fig:distances}, we see the estimated distances for $N=10000$ simulation runs for times $t \in [0, 150]$.
It can be seen that the steady-state behaviour of the system is captured accurately, as the majority of the distances recorded after time $t=60$ lie below the self-distance threshold.
However, the transient behaviour of the system is not captured as accurately.
Upon a more careful inspection of the shape of the histograms in Figure \ref{fig:distributions}, we see that the coarsened system simply converges more quickly to steady-state.
To conclude, we think that the the approximation quality of the steady-state dynamics is a promising result, but a more accurate approximation of the transient behaviour is subject of future work.

\paragraph{Computational Savings}
State-space coarsening results in a more efficient simulation process, since the coarse system is characterised by lower complexity as opposed to the fine system.
We demonstrate these computational savings empirically in terms of the average number of state transitions invoked during simulation.
More specifically, we consider a sample of $5000$ trajectories of the fine and the coarse system.
We have recorded $320\pm25$ initial state transitions on average in each trajectory of the fine system, compared to $56\pm31$ macro-state transitions in trajectories of the coarsened system. The number of transitions in the coarse system is an order of magnitude lower than in the fine one, owing to the reduction of states in the system from a total of 5151 to 10 (the number of macro-states).
\guido{Clearly, our procedure, particularly the GP imputation, incurs some computational overheads.
Table \ref{table:timings} presents the computational savings of using \acp{GP} to estimate satisfaction probability distributions for most states, instead of exhaustively exploring the state-space. All simulations were performed using a Gillespie algorithm implementation, taking 1000 trajectories starting at each examined state, running on 10 cores.}

\begin{table}[h!tbp]
\centering
\begin{tabular}{ c | c | c | c | c}
Sample size &
\begin{minipage}{0.15\columnwidth}
\centering
GP \& MDS time (s)
\end{minipage} &
\begin{minipage}{0.15\columnwidth}\centering
Simulation time (s)
\end{minipage} &
\begin{minipage}{0.15\columnwidth}\centering
Total\\time (s)
\end{minipage}&
\begin{minipage}{0.25\columnwidth}
\centering
Percentage of exhaustive total time (Total time/8516s)
\end{minipage}\\
\hline
100\%&	1616*&	6900&	8516&	100\% \\
50\%&	1133&	3450&	4583&	54\% \\
40\%&	884&	2760&	3644&	43\% \\
30\%&	595&	2070&	2665&	31\% \\
20\%&	354&	1380&	1734&	20\% \\
10\%&	170&	690&	860&	10\% \\
\multicolumn{5}{l}
{\scriptsize * No \ac{GP} was performed here, just the MDS.}
\end{tabular}\newline
\caption{Real running times for simulations of varying sample size (percentage of state-space) and \ac{GP} estimation of remaining states.}
\label{table:timings}
\vspace{-0.6cm}
\end{table}

\section{Discussion}\label{discussion}


We presented a novel approach to the coarsening of a CTMC, in order to gain a stochastic process with a much smaller state-space.  Unlike previous approaches to CTMC aggregation, which are based on structural properties of the state-space, our approach is based on property satisfaction, allowing the coarse-grained system to focus on abstracting the dynamics in terms of aspects of behaviour that are important in the modelling study.  The further steps are to identify key clusters of states in property space, or a lower-dimensional representation of it, and approximate the transition dynamics between them.  For example, this approach might be used within multi-scale modelling to reduce the state-space of a lower level model before embedding in a higher-level representation.  


Common aggregation techniques, such as exact or approximate lumpability, often impose stringent conditions on the symmetries and transition rates within the original state-space. Moreover, the macro-states produced can be difficult to interpret when the reduction is applied directly at the state-space level (i.e.~without a corresponding bisimulation over transition labels). In contrast, the property-based approach allows macro-states to be defined by high-level behaviour, rather than them emerging from an algorithm applied to low-level structure.


The GP regression we employed for estimating satisfaction probability of properties for out-of-sample states proved quite accurate; simulation estimates for 10\% of the states were sufficient to reconstruct the state distribution in the space defined by the probability of property satisfaction, $\phi$-space, without substantial loss of structure.
\guido{Therefore, the proposed approach may be helpful in effectively understanding the behavioural structure of large and complex Markovian systems, with implications for design and verification.}

\guido{Initial experiments on a simple system show that our approach can be practically deployed, with considerable computational savings. The approach induces coarsened dynamics which empirically match the original system's dynamics in terms of steady-state behaviour.} However, the recovery of transient coarse-grained dynamics poses more of a challenge and this will provide a focus for future work.  In particular, we will seek to explore the possibility of quantifying the information lost through the coarsening approach, at least asymptotically, for systems which admit a steady state. \guido{Exploring the scalability of the approach on more complex, higher dimensional examples will also be an important priority. In general, we expect our approach to be beneficial when simulation costs dominate the overheads incurred by the GP regression approach. This condition will be mostly met for systems with moderately large state spaces but complex (e.g. stiff) dynamics. For extremely large state spaces, the cubic complexity (in the number of retained states) of GP regression may force users to adopt excessively sparse sub-sampling schemes, and it may be preferable to replace the GP regression step with alternative schemes with better scalability. Exploration of these computational trade-offs would likely prove insightful for the methodology.}


\bibliographystyle{abbrv} 
\bibliography{CoarseQest}

\end{document}